\documentclass{aa-nov05}
\usepackage{graphicx}
\usepackage{txfonts}

\usepackage{graphicx}
\usepackage{natbib}
\usepackage{subfigure}
\usepackage{ulem}
\usepackage{dcolumn}
\newcolumntype{d}{D{.}{.}{-1}}
\newcommand{\dlabel}[1]{\multicolumn{1}{c}{\mbox{#1}}}
\usepackage{aalongtable}
\usepackage[figuresright]{rotating}

      \def\new#1 {{\bf #1 }}
      \def\cut#1 {\sout{#1} }

\def\kms {$\mathrm{km\,s^{-1}}$} %
\def\AMM {$\mathrm{NH_3}$} %
\def\percc {$\mathrm{cm^{-3}}$} %

\def\ALD {$\mathrm{H_2CO}$} %
\def\ALC {$\mathrm{CH_3OH}$} %
\def\Lsol {$\hbox{L}_\odot$}
\def\Msol {$\hbox{M}_\odot$}

\def\hh {$\hbox{H}_2$}
\begin{document}

   \title{Ammonia in Infrared Dark Clouds}

 %

  \author{T. Pillai
           \inst{1}, 
           F. Wyrowski,
          \inst{1}
           S.J. Carey \inst{2}
          \and
          K.M. Menten  \inst{1}
         }

   \offprints{T.Pillai}

   \institute{Max-Planck-Institut f\"{u}r Radioastronomie, Auf dem H\"{u}gel 69, D-53121 Bonn, Germany \\
              \email{thushara, wyrowski, menten@mpifr-bonn.mpg.de}
               \and
             Spitzer Science Center, California Institute of Technology, MC 314-6, 1200 East California Boulevard, Pasadena, CA 91125\\
              \email{carey@ipac.caltech.edu }}

 \abstract
{While low mass clouds have been
  relatively well studied, our picture of high-mass star formation
  remains unclear. Infrared Dark Clouds appear to be the long sought population of
  cold and dense aggregations with the potential of harbouring the
  earliest stages of massive star formation. Up to now
  there has been no systematic study on the temperature distribution,
  velocity fields, chemical and physical state toward this new cloud
  population.}
   {Knowing these properties is crucial for understanding the
   presence, absence and the very potential of star formation. The
   present paper aims at addressing these questions. We analyse
   temperature structures and velocity fields and gain information on
   their chemical evolution. }
   {We mapped the $(J,K) = (1,1)$ and (2,2) inversion transitions of
   ammonia in 9 infrared dark clouds. Our observations allow the most
   reliable determination of gas temperatures in IRDCs to date.}
 {The gas emission is
remarkably coextensive with the extinction seen at infrared
wavelengths and with the submillimeter dust emission. Our results show
that IRDCs are on average cold ($T < 20 ~ {\rm K}$) and have
variations among the different cores. IRDC cores are in virial
equilibrium, are massive ($M > 100 ~$\Msol), highly turbulent (1 --
3~$ \rm km~s^{-1}$) and exhibit significant velocity structure
(variations around 1 -- 2~$\rm km~s^{-1}$ over the cloud). }
{We find an increasing trend in temperature from IRDCs with high ammonia
column density to high mass protostellar objects and hot
core/Ultracompact H{\sc ii} regions stages of early warm high-mass star
formation. The linewidths of
IRDCs are smaller than those observed in high mass protostellar
objects and hot core/Ultracompact H{\sc ii} regions. On basis of this
sample, and by comparison of the ammonia gas properties within a cloud
and between different clouds, we infer that while active star
formation is not yet pervasive in most IRDCs, local condensations
might collapse in the future or have already begun forming stars.
}


   \maketitle
 \section{Introduction}
 Infrared dark clouds (IRDCs) are cold, dense molecular clouds seen in
 silhouette against the bright diffuse mid-infrared (MIR) emission of
 the Galactic plane. They were discovered during mid-infrared imaging
 surveys with the Infrared Space Observatory (ISO,
 \citealt{perault1996:iso}) and the Mid-course Space Experiment (MSX,
 \citealt{egan1998:irdc}).

In an initial census of a $\sim$180\degr\ long strip of the Galactic
plane (between $269$\degr $< l < 91$\degr, $\rm b = \pm 0.5$\degr),
\citet{egan1998:irdc} found $\sim$2000 compact objects seen in
absorption against bright mid-infrared emission from the Galactic
plane. Examination of 2MASS, MSX and IRAS images of these objects
reveals that they appear as shadows at all these wavelengths, although
they are best identified in the $8.3 \mu$m MSX band, because,
first, the 7.7 and 8.6~$ \mu$m PAH features associated with
star-forming regions contribute to a brighter background emission and,
second, the MSX $8.3 \mu$m band is more sensitive than the satellite's other
bands. Recently, we have reviewed the observational studies on IRDCs
(\citealt*{menten2005:iau}).

While low mass clouds have been relatively well studied, our picture
of high-mass star formation %
remains unclear \citep[see][]{evans2002:aspc}.  IRDCs appear to be the
long sought population of cold and dense aggregations with the
potential of harbouring the earliest stages of massive star
formation. It is likely that some of the stars forming in them are
massive (luminosities of submm condensations range up to 10$^{4}$
\Lsol). A recent study by \citet{ormel2005:irdc} on an IRDC towards
the W51 Giant Molecular Cloud (GMC) suggests that sources of $\sim
300$~\Msol\ are embedded within the cores, most likely protostars.
Recently, we \citep{pillai2005a:g11} reported a detailed study of the
strongest submm peak in the IRDC G11.11$-$0.12, where we find clear
evidence of a heavily embedded protostar.

The salient results on IRDCs are summarised below.  The IRDCs observed
so far have sizes of 1-10~pc and have mostly a filamentary
morphology. On the basis of LVG calculations of mm \ALD\ observations
\citet{carey1998:irdc} find that typical IRDCs have gas densities of
$n > 10^{6} ~ {\rm cm}^{-3}$ and temperatures of $T < 20 ~ {\rm
K}$. Kinematic distances determined from \ALD\ observations using a
standard Galactic rotation curve that ranges between 2.2 and 4.8~kpc
(\citealt{carey1998:irdc})
indicate that the clouds are
not local. All observed IRDCs in the sample of \citet{carey2000:irdc}
contain 1 -- 4 bright sub-millimeter (submm) dust continuum emission
peaks ($>$ 1 Jy/ 14 \arcsec\ beam at 850 $\mu$m) surrounded by an
envelope of emission which matches the morphology of the IRDC in
mid-infrared extinction. The cores corresponding to the brightest
submm peaks have masses 100 and 1200 \Msol, except for two
clouds in the Cygnus region that have masses around 40 \Msol.

\citet{hennebelle2001:irdc} in a systematic analysis of the ISOGAL
images extracted about 450 IRDCs, for which they derive $15 \mu$m
optical depths of 1 to 4. \citet{teyssier2002:irdc} reported that Large
Velocity Gradient (LVG) model calculations of HC$_3$N, $^{13}$CO, and
C$^{18}$O yield densities larger than $10^5$ \percc\ in the densest
parts. The authors claim to find kinetic temperatures between 8 and 25
K based on CH$_3$CCH observations; the higher values being found
toward embedded objects, however a detailed analysis is hitherto
unpublished.

Observations of molecules in IRDCs so far have concentrated on just a
few species.  Up to now there has been no systematic study on the
temperature distribution, velocity fields, chemical and physical state
toward this new cloud population.  Knowing these properties is crucial
for understanding the presence, absence and the very potential of star
formation.  The present paper aims at addressing these questions.  We
have started with a survey of a sample of 9 IRDCs in (1,1) and (2,2)
cm rotational transitions of ammonia (\AMM). This sample
has been studied before by
\citet{carey1998:irdc} and \citet{carey2000:irdc} and was selected
on the basis of the large extent and high contrast of the IRDCs against the MIR
background.

Ammonia has proven to be an important tool in measuring the physical
conditions in molecular clouds (\citealt{ho1983:nh3}). Since only the
lowest \AMM\ energy levels are expected to be populated in cool dark
clouds  ($T<20$~K), their physical conditions  can be
probed using the (1,1) and (2,2) inversion transitions in the
metastable  ($J,K$) rotational levels of ammonia. Radiative transitions
between different $K$-ladders are forbidden, therefore the lowest levels
are populated only via collisions. The optical depth can be
determined from the ratio of the hyperfine satellites. Thus, the
population of the different levels can be estimated and hence the
temperature of the gas determined. In addition, recent chemical models
reveal that \AMM\ (and also N$_2$H$^+$), does not deplete from the gas
phase for the densities observed in IRDCs ($<10^{6}$~\percc)
\citep{bergin1997:chemistry}. Thus \AMM\ is an excellent tracer of the
dense gas where many other molecules would have heavily depleted.

In \S\ref{sec:observations}, we give details of our observations with
the Effelsberg 100m telescope.
In \S\ref{sec:results}, we discuss the
data reduction and present the correlation between gas emission and
MIR absorption. In \S\ref{sec:analysis}, we derive the
rotational temperature,
gas kinetic temperature and \AMM\
column density. Furthermore, we analyse the velocity structure,
estimate dust mass, virial mass and \AMM\ abundance. In
\S\ref{sec:disc}, we compare the core gas properties (temperature,
\AMM\ linewidths and column density) with that of other populations of
objects that are thought
to trace the early stages of high mass star
formation. We do a similar comparison with %
local dark
clouds well-studied in \AMM. Finally, we speculate on a possible
formation mechanism of IRDCs involving  supernova remnants.

\section{Observations \label{sec:observations}}

 We mapped the IRDCs listed in Table.~\ref{tab:source_list} with the
 Effelsberg 100m telescope of the Max-Planck-Institut f\"ur
 Radioastronomie in October 1999.  The frontend was the facility 1.3
 cm maser receiver tuned to a frequency of \mbox{23.7 GHz} centered
 between the \AMM\ (1,1) and (2,2) transitions.  The spectrometer was
 a 8192~channel auto-correlator used with 2 subunits of
 20~MHz bandwidth each. The resulting spectral resolution was $\approx
 0.2$~km$\,$s$^{-1}$ after smoothing the data to increase the
 signal-to-noise ratio. The beamwidth at the frequencies of the \AMM\
 lines is 40\arcsec\ (FWHM). The observations were conducted in
frequency switching mode with a frequency throw of 7.5~MHz. The maps
 toward all sources covered the extinction seen in the MSX images and
 were made with half power beamwidth (HPBW) spacing (40\arcsec). Alternate scans
were inter-spaced at half the full beamwidth resulting in $\approx
20\times\sqrt(2)$ spacing.  Pointing was
 checked at roughly hourly intervals by means of continuum drift scans
 on nearby pointing sources. We found the pointing to be accurate to
 within 12\arcsec.

Absolute calibration is not an issue in determining the rotational
temperature, because it is solely governed by the ratio of the
\AMM(1,1) hyperfine lines and the (1,1) and (2,2) brightness
temperatures.  However, in order to estimate the excitation
temperature and the column density the data needs to be
calibrated. The calibration procedure is documented on our webpage
\footnote{www.mpifr-bonn.mpg.de/staff/tpillai/eff\_{}calib/index.html.}. The
important steps are mentioned below.  The 100m data (normally in CLASS
format) is in arbitrary noise tube units ($T_{\rm N}$) and has to be
converted to main beam brightness temperature units. We observe a
standard flux calibrator with known flux $S_{\lambda}$ at wavelength 
$\lambda$ and thus estimate the conversion factor from $T_{N}$ scale to 
$\rm Jy$.

The main beam brightness temperature $T_{\rm MB}$ of the calibrator
for a given beamwidth $\theta$ and wave length $\lambda$ is then
given by
\begin{equation}
     (T_{\rm MB}/\rm K)= \frac{(S_{\rm \lambda}/\rm Jy)
     \times(\lambda/cm)^2}{(\theta/arcmin)^2 \times 2.65}.
      \end{equation}

The primary flux calibration is based upon continuum scans of NGC7027
assuming a flux density of $\approx 5.6$~Jy
(\citealt{ott1994:Fluxes}). The Ultracompact H{\sc ii} (UCH{\sc ii})
regions G10.62-0.38 and G34.26+0.15 and the quasar J1743-038 were used
as secondary calibrators which span the entire elevation range and an
elevation-dependant calibration factor was derived. The maximum
rms uncertainty in calibration for elevations lower than
$19^{\circ}$ is 15\%.

\section{Data Reduction and Analysis \label{sec:results} }

The spectra were reduced using the CLASS package
\citep*{forveille1989:class}.  For a given source, the spectra were
averaged and a polynomial baseline of order 3 -- 5 subtracted.

The maps 
were generated by convolving the original data using a
Gaussian function with a HPBW of 40\arcsec. 
Table~\ref{tab:source_list} reports the coordinates of the central
position for each map.

\begin{figure*}
\begin{center}
\includegraphics[height=\linewidth]{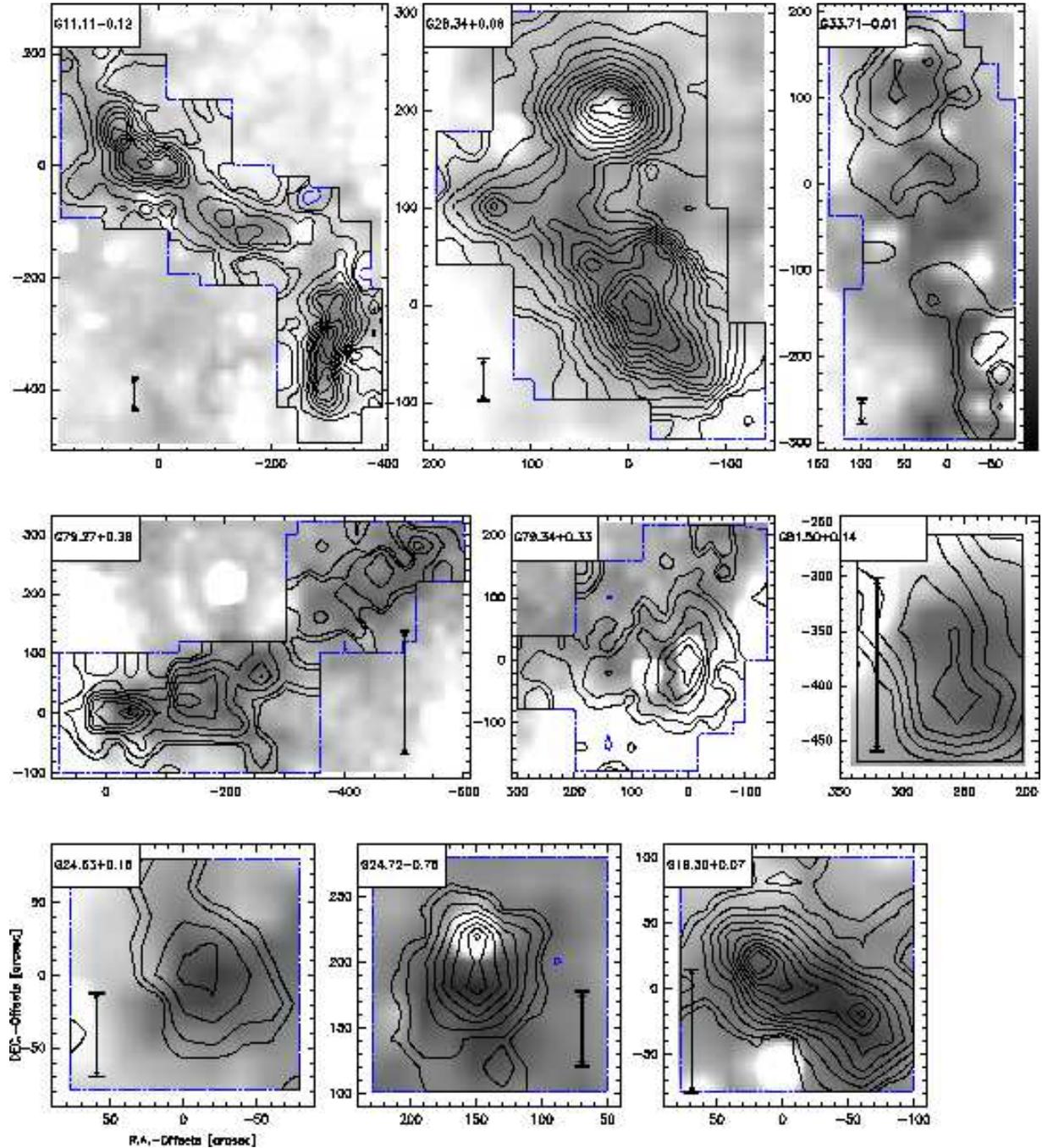}
\end{center}
\caption{ MSX image of the clouds at 8.3${\mu}$m (greyscale) with \AMM\
(1,1) integrated intensity as contours. The \AMM\ (1,1) maps have a
resolution of 40\arcsec and the same spacing.  The greyscale
corresponds to intensity range shown by the wedge in the upper right
corner (from dark to light). The contour levels are $-2\sigma$,
 $1\sigma$, $2\sigma$, $4\sigma$, $6\sigma$... The rms
noise levels are given in Table~\ref{tab:line_parameters}.  The
approximate map boundary is also indicated. Tick marks are coordinate
offsets (in arcseconds) relative to the positions given in
Table~\ref{tab:source_list} .  The bar indicates a length of 1 pc at the
distance of each IRDC.
\label{fig:msx_nh3}}
\end{figure*}

MSX 8${\mu}$m images of these clouds with superimposed contours of the
\AMM\ (1,1) integrated intensity are shown in Fig.~\ref{fig:msx_nh3}.
The velocity range used for
integration is within $\pm 25$~ \kms of the 'Local Standard of Rest
Velocity', $V_{\rm LSR}$, hence including the satellite lines.  The
\AMM\ maps of G79.27+0.38 and G33.71-0.01 have clumpy structures.  Some of the
clumps may be artificial and caused by the low signal-to-noise ratio.

\begin{figure*}
\begin{center}
\includegraphics[height=\linewidth]{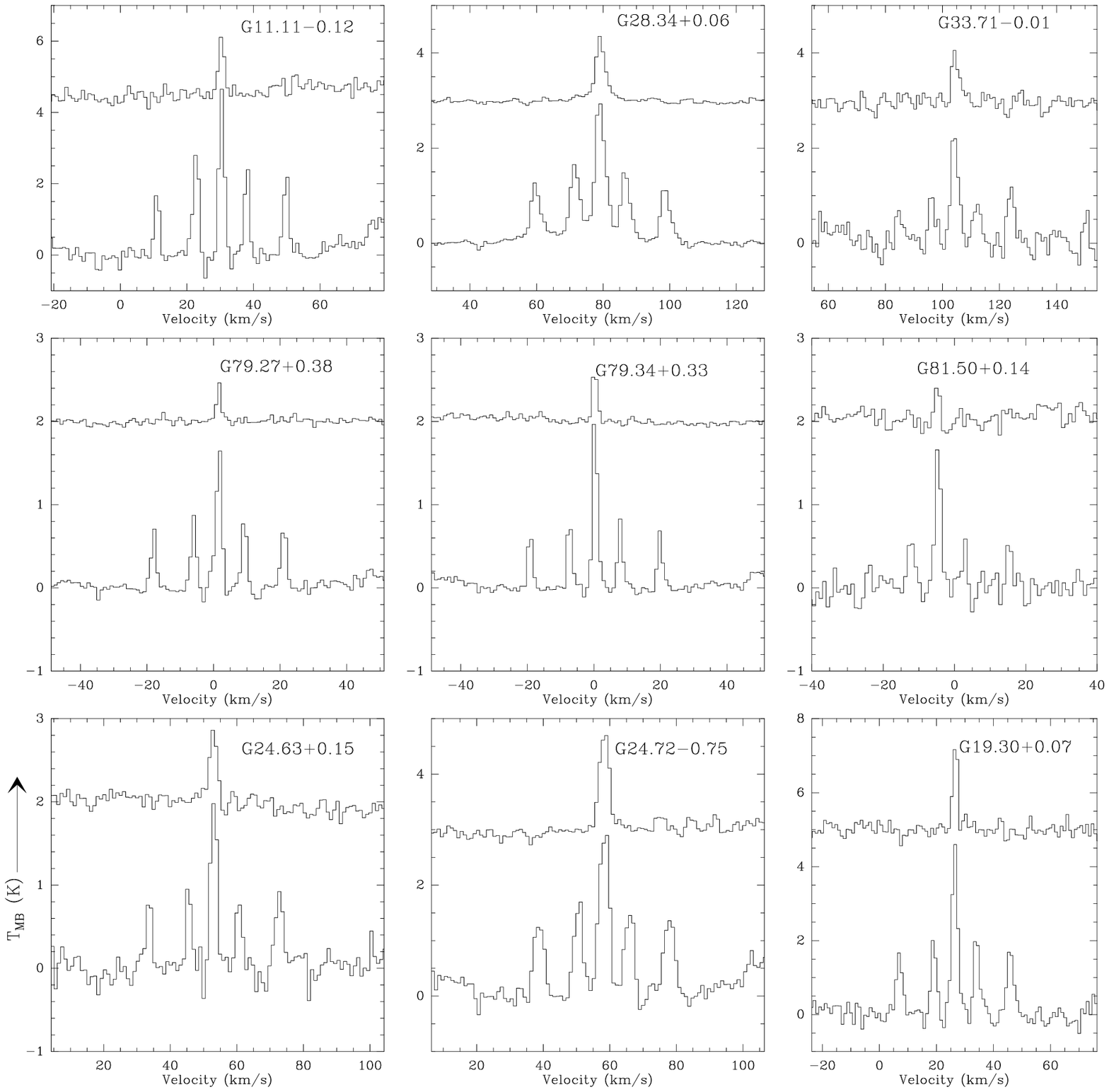}
\end{center}
 \caption{ \AMM(1,1) (lower histogram in each panel) and (2,2) (upper
 histogram, offset from 0 K for clarity) spectra toward the brightest position of the 9 observed
 sources. The brightest positions are referred to as \AMM\ P1 in
 Table~\ref{tab:physical_parameters}\label{fig:nh3_spec}.}

\end{figure*}

The reduced (1,1) and (2,2) spectra for the \AMM\ peak positions of
the observed sources are shown in Fig.~\ref{fig:nh3_spec}. \AMM(1,1)
observations were reduced using ``METHOD NH3(1,1)'' in CLASS to fit
the hyperfine structure and derive optical depths and linewidths. The
standard procedure to analyse the \AMM\ (1,1) and (2,2) lines have
been described in detail by \citet{bachiller1997:nh3}.

The hyperfine structure of the (2,2) line is too weak to be observed;
therefore, the optical depth could be obtained only for \AMM(1,1). The
\AMM(1,1) and (2,2) main beam brightness temperatures are obtained by fitting the main line with a single
Gaussian. The (1,1) and (2,2) linewidth is obtained by the hyperfine fitting
which accounts for the line broadening due to optical
depth effects excluding self-absorption.
The parameters obtained from the fits to the (1,1) hyperfine lines are
listed in Table~\ref{tab:line_parameters}.

\label{fig:2}

\label{fig:3}

The basic physical parameters, namely the excitation temperature,
rotational temperature, the kinetic temperature and ammonia column
density, have been derived using the standard formulation for \AMM\
spectra (\citealt{ho1983:nh3}). Table~\ref{tab:physical_parameters}
summarises the estimates of these parameters toward the cores. We give
the formal errors (1$\sigma$), derived from Gaussian error
propagation. 
\section{Results \label{sec:analysis}}
Our sample consists of both 'Infrared dark clouds', with extents of
$\sim$1--10~pc, and 'Infrared dark cloud complexes', which are comprised of
multiple individual clouds.
IRDC G11.11-0.12 would thus be a complex ($> 4$~pc) while G24.63+0.15
would be a dark cloud. The IRDCs G79.34+0.33 and G79.27+0.38 are
essentially two parts of the same extended filament in the Cygnus-X
region connected by a bright patch of dust emission as seen in SCUBA
and MSX images \citep{redman2003:g79}.

\subsection{Cloud Morphology \label{subsec:morph}}

There is generally a close match
between the ammonia emission and the mid-infrared extinction as shown in Fig.~\ref{fig:msx_nh3}. In
G24.72$-$0.75 and G79.34+0.33, the \AMM\ emission peak is however
correlated with bright and compact MIR
emission. \citet{redman2003:g79} find that the MIR emission in
G79.34+0.33 corresponds to a luminous 'Young Stellar Object' (YSO)
with a strong IR excess which might be interacting with the foreground
IRDC. No such interaction has been reported for the IRDC G24.72$-$0.75
and the nature of the MIR object is unknown.

\begin{figure}
\begin{center}
\includegraphics[height=\linewidth, angle=-90]{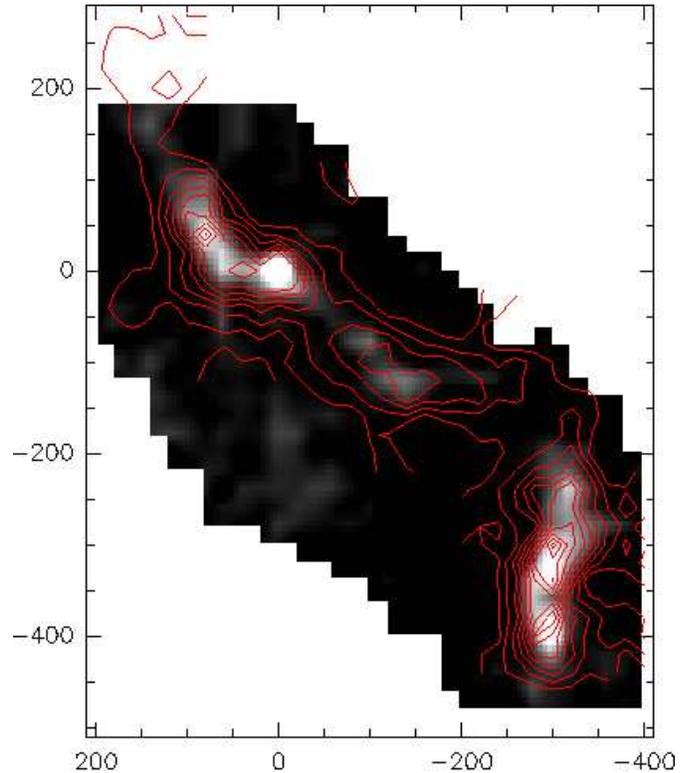}
\end{center}
\caption{SCUBA 850${\mu}$m image \citep{carey2000:irdc} of the
 cloud G11.11-0.12 with \AMM\ (1,1) integrated intensity as
 contours. The contour levels are multiples of $2\sigma$. 
SCUBA image has a resolution of 14\arcsec. \label{fig:g11_scuba}}

\end{figure}

In general, the cloud geometry is extended and filamentary and in no case
close to spherical.  The mean aspect ratio determined by fitting 2D
ellipse to the entire \AMM\ emitting region is 2.2 and the total
extent of the clouds ranges from 0.4 -- 8.3~pc. G11.11-0.12 is a
filamentary cloud as revealed by the 8${\mu}$m extinction and the
850${\mu}$m dust emission as shown in Fig.~\ref{fig:g11_scuba}. 
 The peaks of the submm emission are
strikingly coincident with the ammonia cores. The \AMM\ map reveals at
least two peaks toward the north-east segment and another set of
peaks towards the southwest extension of the filament suggesting
that there are several unresolved sub-structures or cores within the extended
filament.

\begin{figure}
\begin{center}
\includegraphics[height=\linewidth, angle=-90]{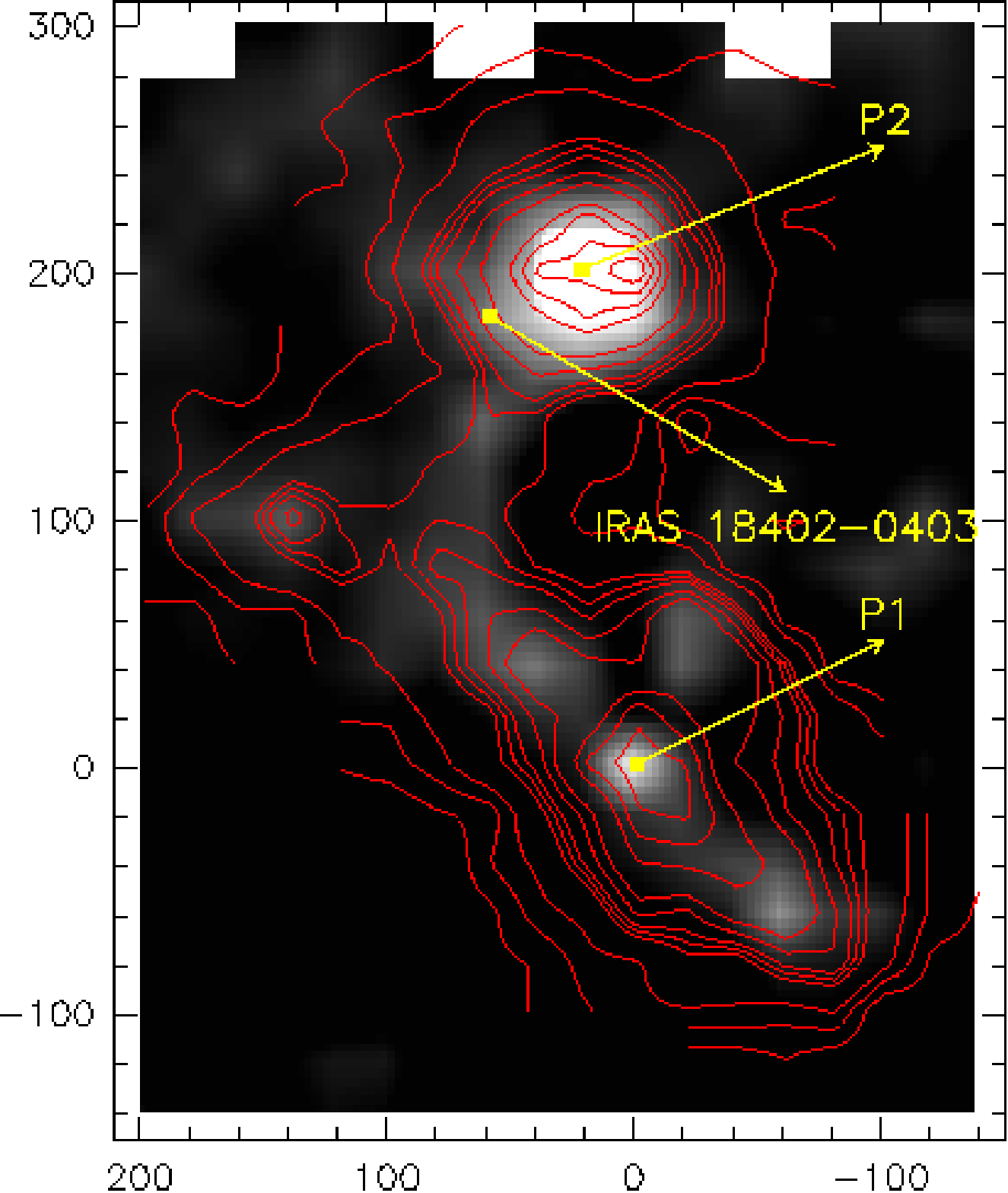}
\end{center}
\caption{ SCUBA 850${\mu}$m image of G28.34 with \AMM\ (1,1)
 integrated intensity as contours. The contour levels are ($2\sigma$, $4\sigma$, $6\sigma$...).  \label{fig:g28_scuba}}
\end{figure}

 The strongest submm emission peak P2 in
G28.34 is in
the close vicinity of the IRAS source 18402-0403. But the peak of the
ammonia emission for the northern extension seems to be offset from
that of the IRAS source (see Fig.~\ref{fig:g28_scuba}).
 This could be due to the interaction of the
IRAS source with the cloud.

\label{fig:4}

\label{fig:5}
\subsection{Source Sizes \label{subsec:size}}

Due to the large distances to the clouds, the compact structures
within the cloud are marginally resolved with the 40\arcsec\ beam of our
\AMM\ observations. Hence, estimating the size of the core from \AMM\ might only
deliver upper limits of the order of the beam size. We use channel
maps to solve the problem of identifying clumps. The information in
the third dimension enables us to disentangle clumps that are
sufficiently well separated in velocity. For those sources with SCUBA
850${\mu}$m data, we cross-identify the clumps within half
the \AMM\ beam ($\sim 20$~\arcsec) at the higher SCUBA resolution ($\sim
14$~\arcsec) and estimate the source size by a 2D Gaussian fit to the
clumps in the SCUBA data. In cases where we do not have dust emission
maps, we use the 2D Gaussian fit routine in GRAPHIC. It searches for the
brightest pixel in each velocity channel of the channel map across the
\AMM\ (1,1) main component and fits a 2D Gaussian to determine source
size. The output is then cross-checked over the different channels and
the fit obtained for the brightest emission is used. We find 1 -- 5
clumps for each cloud. The source sizes after correcting for the beam
size (after subtracting the Gaussian beam size in quadrature) are listed in
Table~\ref{tab:physical_parameters}. The dense cores within these
IRDCs are thus not resolved with the  $40~\arcsec$ 100m beam.
We find several secondary peaks offset by $>$~1$\sigma$ in most
sources; offsets from the central position are given in
Table~\ref{tab:physical_parameters}. Table~\ref{tab:physical_parameters}
should be referred to for the nomenclature of the clumps identified in
the \AMM\ maps and/or the SCUBA 850~$\mu$m images hereafter.

\subsection{Line Profiles}

In general, \AMM\ lines are brighter than the mm rotational
\ALD\ transitions observed  by \citet{carey1998:irdc}. As given in
Table~\ref{tab:line_parameters}, the (2,2) peak line intensity is on
average 40\% of that of the (1,1) line. Gaussian line profiles with extended
wings have been reported for \ALD\ lines in most of the clouds
\citep{carey1998:irdc} but we do not observe any pronounced
wings in \AMM. Leurini et al.\ 2006 (submitted) find line profiles similar to those
observed in \ALD\ in the mm
transitions of CH$_3$OH.

\subsection{Linewidth \label{subsec:linewidths}}

 The width of a spectral line is an important parameter because it is a measure of
 the total kinetic energy of the cloud. The \AMM(1,1) linewidths for
 all sources lie between 0.8 -- 3 ~$\rm km~s^{-1}$. The \ALD\
 linewidths are significantly larger (factor 2) than the \AMM\
 linewidths in all cases except G81.50+0.14, where the \AMM\
 linewidths are twice as large.

The \AMM\ linewidths for our sample are higher than those of
\AMM\ cores reported in \citet{jijina1999:database}, which are mostly
low mass cores. The large linewidths might be explained as due to
clumping. Clumps with smaller linewidth but with a higher
clump-to-clump velocity dispersion may add up to the observed linewidths. 
The sources closest to us (G79 IRDCs) have the smallest linewidth and belong 
to the Cygnus-X region. In order to test whether the
larger linewidth we observe at larger distance is a distance effect,
we average the G79 IRDC linewidth over an area $\delta A \propto (d_{\rm
far}/d_{\rm Cyg})^2$ where ${d_{\rm far}}$ is the distance to a source
farther than Cyg and with a much higher linewidth. On comparing the
resultant scaled values, we find that the linewidths are still
significantly larger for the sources at larger distances after accounting for
the larger region in each beam for the more distant cores.

To illustrate this point, let us take the example of the two
extreme values of linewidths from our sample. Dark core G79.27+0.38
\AMM\ P2 (at 1~kpc) has a linewidth at the brightest position of
0.83~$\rm km~s^{-1}$ while G28.34+0.06 \AMM\ P2 (at 4.8~kpc) has a
linewidth of 2.65~$\rm km~s^{-1}$. Neither of the two cores have a
MIR counterpart which might indicate any deeply embedded protostar
influencing its immediate environments.  Averaging all the
emission in G79.27+0.38 over an area of $\sim 200 \times 200\arcsec$, we 
obtain a linewidth of 1.5~$\rm km~s^{-1}$, still smaller than
2.65~$\rm km~s^{-1}$. Therefore, the large linewidths derived appears
to be mainly due to the velocity dispersions within the beam.

One of the assumptions made to derive the temperature is that the beam
filling factor is the same for both inversion lines; that is, 
the two lines are emitted by the same volume of gas. The (2,2) linewidths 
are slightly larger than the (1,1) linewidths 
(cf. Table~\ref{tab:line_parameters}) toward only some cores,
suggesting that the same gas is not exactly traced by both lines in those 
sources.

\begin{figure}
\begin{center}
\includegraphics[height=1.2\linewidth, angle=0]{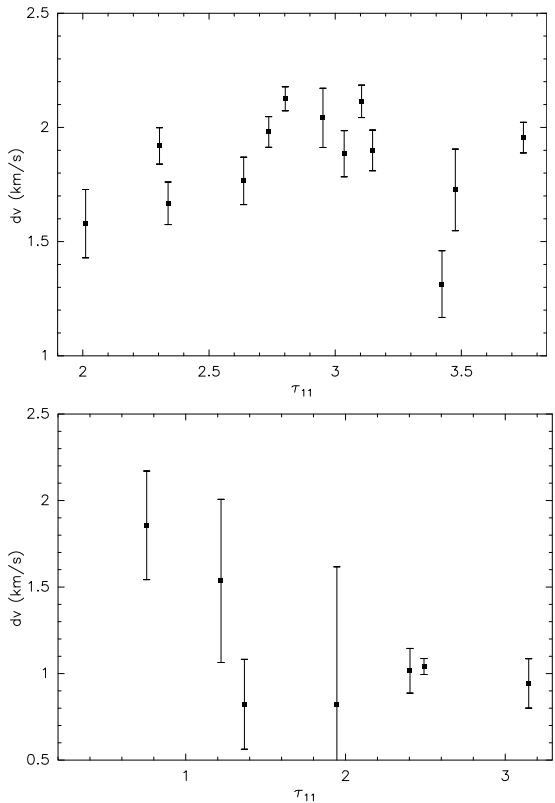}
\end{center}
\caption{Correlation plot between the \AMM(1,1) optical depth
and the \AMM\ (1,1) linewidth. \textit{Bottom panel}: G79.27$+$0.38 \AMM\ P1, \textit{Top panel}: G28.34+0.06 \AMM\ P2. \label{fig:linewidth_correln}}

\end{figure}

The linewidths
exhibited by these sources are much larger than the thermal linewidth,
$\Delta v_{\rm th}$, which
for $T_{\rm kin} \sim 20 K$ should be $\sim 0.22
~ \rm km~s^{-1}$ as per the relation
\begin{equation}
        \Delta v_{\rm th} \sim \sqrt{ \frac{8 \ln(2) k \, T_{\rm
        kin}}{m_{\rm NH_3}}},
\end{equation}
where $k$ is Boltzmann's constant and $m_{\rm NH_3}$ is the mass of the
ammonia molecule.
 Average linewidths $\sim 2 ~ \rm km~s^{-1}$ could be explained in
 terms of velocity gradients %
 due to rotation of the cloud,
 or turbulent cloud movements. The linewidths seems to be especially
 high towards the cloud G28.34+0.06, where there are several
 unresolved \AMM\ clumps, identified in the SCUBA images. In the G79
IRDC the linewidth decreases from G79.34$+$0.33 \AMM\ P1
 ($1.22\pm0.03$~\kms) to G79.27$+$0.38 \AMM\
 P3($0.82\pm0.03$~\kms). As shown in Fig.~\ref{fig:linewidth_correln},
 we find that the linewidth anti-correlates with optical depth for G28.34
 while there is a possible positive correlation G79.27. 

G79.27+0.38 \AMM\ P1 seems to be more quiescent than its
surroundings.  In the G79 cloud complex, there is considerable
difference in linewidth ($\sim 0.5$~\kms) between the different
clumps along the cloud from east to west. Thus, these clouds might be
harbouring objects at different stages of evolution \citep{redman2003:g79}.
\label{fig:6}
\subsection{Kinetic Temperature \label{subsec:tkin} }

 Since the optical depth is known, we derive the excitation
 temperature of the \AMM\ (1,1) inversion transition
 (Table~\ref{tab:physical_parameters}) via the relation

\begin{equation}
        T_{\rm EX} = {\frac{T_{\rm MB}}{1-\rm e^{-\tau}}}+2.7~{\rm K},
\end{equation}
 where $T_{\rm MB}$ and $\tau$ represents the temperature
 and the optical depth of the of the (1,1) line.  By
 fitting the main and the hyperfine components of the (1,1) line and
 the main component of the (2,2) line, we obtain the rotational
 temperature. An analytical expression
 \citep[][]{tafalla2004:starless_cores} has been used to estimate
 the kinetic temperature from the rotational temperature.
The kinetic temperature is given by the expression

\begin{equation}
        T_{\rm kin} = {\frac{T_{\rm rot}}{1-\frac{T_{\rm rot}}{42}\rm ln[1+1.1exp(-16/T_{\rm rot})]}},
\end{equation}

The typical temperatures for the cores range from 11 -- 17~K
as displayed in Table~\ref{tab:physical_parameters}. In some cases, we
find that the fit to the \AMM(1,1) line  slightly underestimates the peak
intensity.

We find that the kinetic temperature is significantly higher than the
excitation temperature, as given in
Table~\ref{tab:physical_parameters}. The beam filling factor $\eta$ is
a measure of the fraction of the beam filled by the observed source. Assuming
that the cores are in local thermodynamic equlibrium (LTE), we
may estimate this fraction as $\eta = T_{\rm EX}/T_{\rm kin}$ where
$T_{\rm Kin}$ is the kinetic temperature of the gas.  We derive filling 
factors of $\sim0.3$ -- 0.5 for all the clumps. These low filling factors
suggest either sub-thermal excitation (non LTE conditions) or
clumping within the beam. From our estimation of the sizes
(Table~\ref{tab:physical_parameters}), we find most of the cores are
unresolved with the 40$''$ beam. Thus clumping is more likely to
explain the small filling factors.

We find hints of temperature gradients (inside-out) within the
cores in three sources and a reverse gradient in one source. However the large
error bars associated with the rotational temperature does not allow
us to make a convincing case. The temperature structure will be analysed
in detail in a future paper with observations at high angular resolution (Pillai et al.\ 2006 in prep).

There is also a variation in temperature within the different cores of
the same cloud.  In G11.11-0.12, with a projected extent of a few pc,
the average (over the core size) temperatures of the
southern cores are 2 -- 3~K ($10.9\pm1$~K) lower than the northern
cores ($13.4\pm1$~K).

In G28.34+0.06, the gas temperature derived towards \AMM\ P3,
$12.7\pm1$~K and \AMM\ P5, $13.6\pm1$~K is significantly lower than in
the rest of the cloud. \ALC\ observations on the P1 position (Leurini
et al.\ 2006 submitted) reveal a cold and a hot component, the latter
with a small filling factor.  We obtain kinetic temperatures of $\sim
16~ \rm K$ towards the peaks of \AMM\ P1 and \AMM\ P2.  Higher angular
resolution observations are needed to confirm a hot component of small
extent also in ammonia.

In G19.30$+$0.07, the temperature
at the position of peak \AMM\ emission P1 is $17.9\pm1.6$~K, much warmer than the rest of the
cloud ($13.1\pm1.0$~K).

In the G79 complex, we find that the temperature in G79.34$+$0.33 \AMM\ P1
is higher
($14.3\pm0.8$~K) than the other 2 cores G79.27$+$0.38 \AMM\ P1
($12.2\pm0.8$~K) \& G79.27$+$0.38 \AMM\ P2 ($11.1\pm1.2$~K) in the
western part of the filament.

\subsection{Velocity Structure \label{subsec:velocity} }

\begin{figure}
\begin{center}
\includegraphics[height=\linewidth, angle=-90]{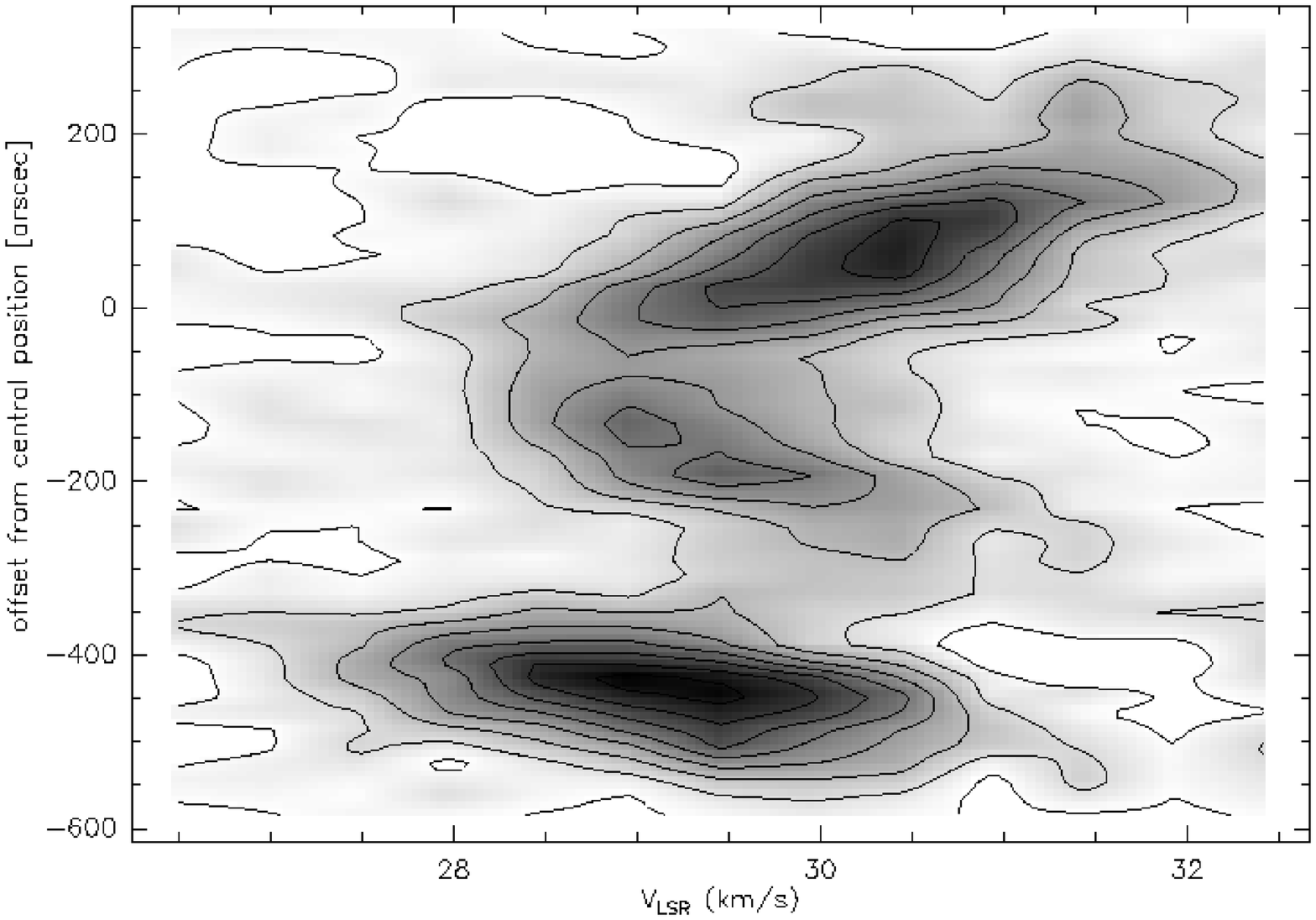}
\end{center}
\caption{The average velocity-position map of \AMM\ (1,1) for the cloud G11.11-0.12  at a position angle of $\sim 40$~\degr. Note that the Y axis is
the offset from the reference position, given in
Table~\ref{tab:source_list}. offsets run from south-west (negative
offsets) to north-east (positive offsets). \label{fig:g11pv}}
\end{figure}

\begin{figure*}
\begin{center}
\includegraphics[height=\linewidth, angle=-90]{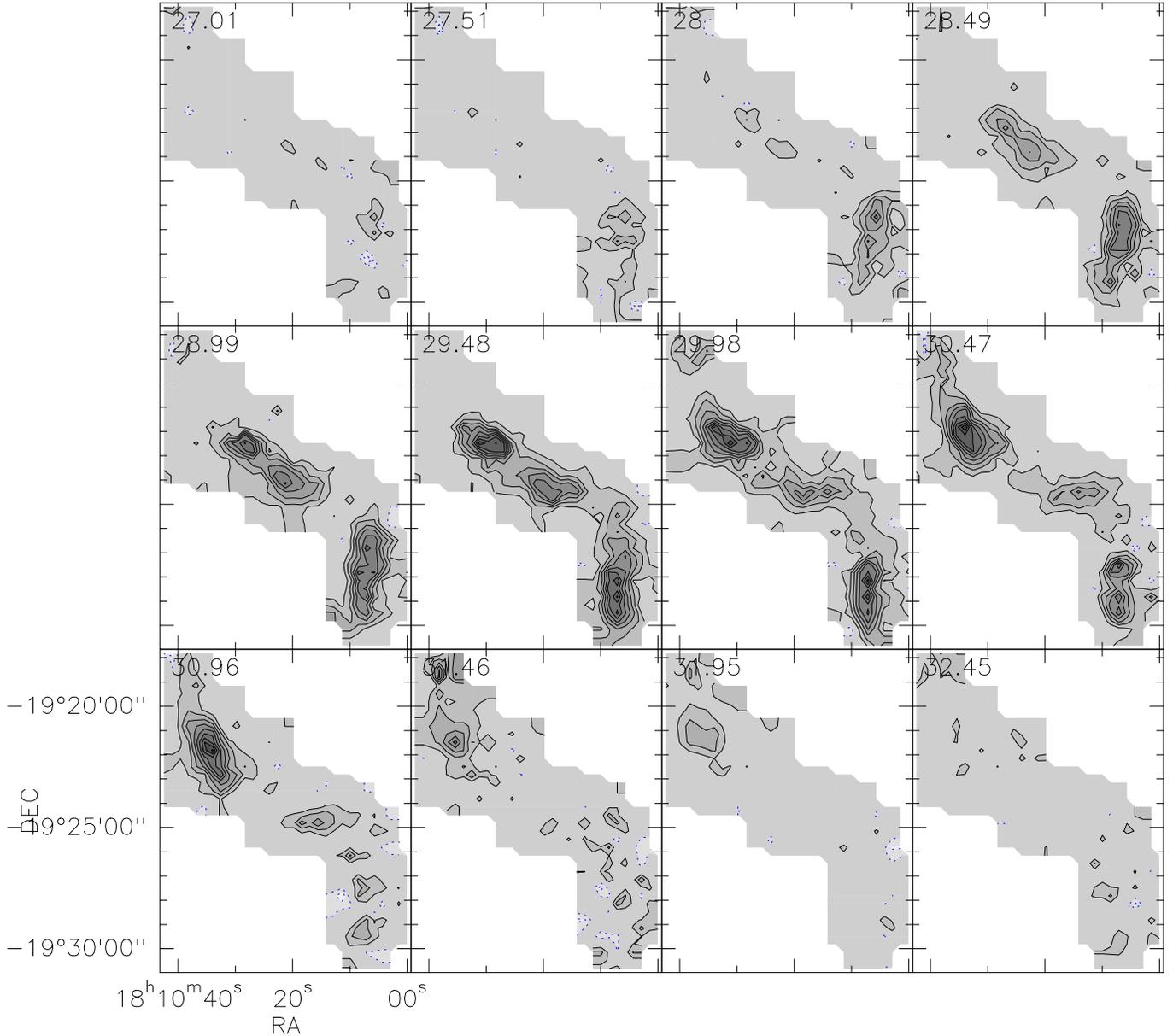}
\end{center}
\caption{Channel maps for G11.11$-$0.12 over the main component
for the \AMM(1,1) transition. \label{fig:g11chan}}
\end{figure*}

The average velocity-position diagram along an axis with position
angle of $+40~$\degr\ (to align the position axis roughly with the direction
of the filament on the sky) for the cloud G11.11-0.12 is shown in
Fig.~\ref{fig:g11pv}. At every position along the position axis, the
\AMM\ (1,1) spectra were averaged along a line perpendicular to the
position axis.
There is a clear trend for the velocities to decrease from the south
to the P1 position and increase to the north of the NE filament. This
is also seen in the channel maps shown in Fig.~\ref{fig:g11chan} where
the different clumps in the north and south appear at distinctly
separate velocities. Given that the \AMM\ (1,1) linewidth at all
positions is $\le 2.5$ \kms , this velocity shift ($>$ 5~\kms ) between the peaks
is certainly significant. The unresolved clump
G11.11$-$0.12 \AMM\ P4 is more associated with the G11.11$-$0.12 \AMM\
P2 clump while P3 fades away at those velocities.

Recently, this filament had been studied in absorption against the
 diffuse 8 $\mu$m Galactic background and in emission from cold dust
 at 850 $\mu$m \citep{johnstone2003:g11}. They model the 850 $\mu$m
 emission by fitting a non-magnetic isothermal cylinder profile to the
 radial structure of the entire filament (south and north segment).The
 velocity structure observed in \AMM\ (1,1), however, shows several distinct
clumps. This suggests that the density structure of the filament cannot
be rigorously described by a simple continuous cylinder.

\label{fig:8}

\label{fig:9}

\begin{figure*}
\begin{center}
\includegraphics[height=\linewidth, angle=-90]{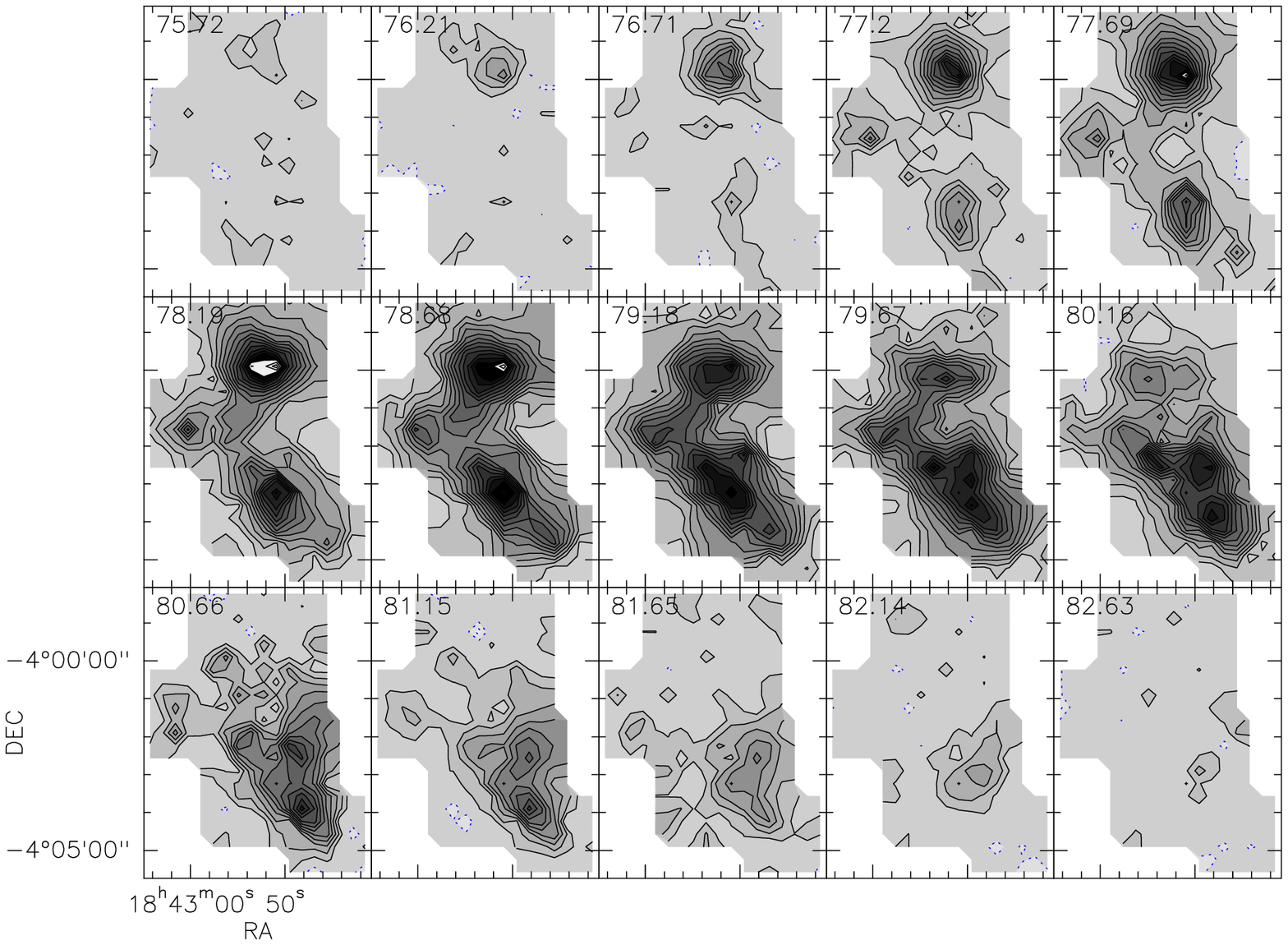}
\end{center}
\caption{channel map for G28.34+0.06 over the main component for the \AMM(1,1) transition.\label{fig:g28chan} }
\end{figure*}

The channel maps for G28.34+0.06 are shown in Fig.~\ref{fig:g28chan}. The two
extensions of the cloud around P1 and P2 differ in velocity by about
1.5~\kms. This is the difference in velocity for the two peaks at
which they are brightest.  From the channel maps it appears that there
is a bridge between the two parts of the cloud connecting P1 and
P2. At the velocity of 80.16~\kms, the three unresolved clumps, which are
identified in dust continuum emission but not clearly identified 
in the \AMM\ (1,1) integrated intensity emission, are revealed.

\label{fig:10}

\begin{figure*}
\begin{center}
\includegraphics[width=\linewidth]{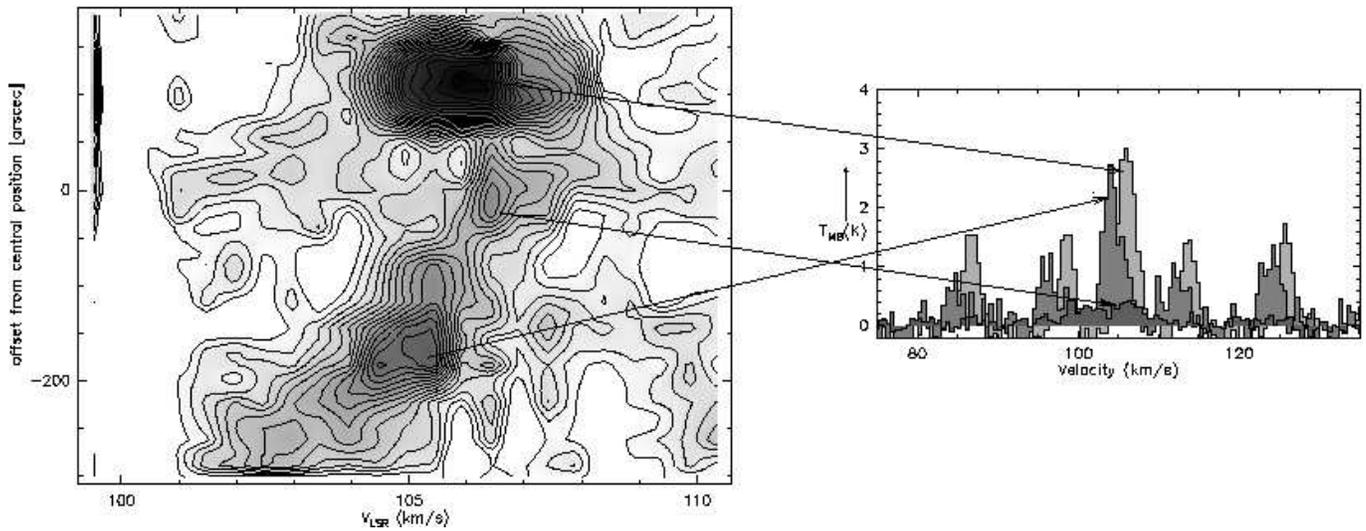}
\end{center}
\caption{\textit{Left panel:} The average velocity-position map of \AMM\ (1,1)
along the declination axis for the
cloud G33.71$-$0.01. \textit{Right panel:} The spectra obtained at the
corresponding position in the map (indicated by the arrows). Note the
large linewidth component toward the centre part of the cloud and the change in
LSR velocity from south to north.\label{fig:g33pv} }
\end{figure*}

G33.71$-$0.01 shows another interesting case of distinct velocity
variations across a cloud. The average velocity position map along
the declination axis is displayed in Fig.~\ref{fig:g33pv}. The velocity
increases from the south at $\sim103$~\kms towards the north to
$\sim106$~\kms\ while showing a very wide weak component at
$\sim105$~\kms\ at the (0,0) position. Its detection at (0,0) is only
at the 2$\sigma$ level but is also observed in \AMM\ (2,2). A possible
interaction with the nearby SNR G33.6+0.1 is discussed in \S5.2.

\label{fig:11}

\subsection{Column Density, \AMM\ Abundance and Masses \label{subsec:masses} }

 We have estimated the column densities and total masses for the
 bright, compact sources for which the SCUBA $850 ~ \rm \mu $m maps
 are available, after smoothing the data to the resolution of 100m beam (40\arcsec). 
Assuming a dust opacity of $\kappa_{\rm m}$ of
 1.85 $\rm cm^2/g$ at $850 ~ \rm \mu $m for grains with thick ice
 mantles and gas density $\rm n(H)= 10^6$\percc \citep{ossenkopf1994:opacities}, 
the effective $\rm H_2$ column density is \citep{launhardt1996:thesis}

\begin{equation}
      N({\rm H_2}) = {\frac{6.2\times10^{16}S_{\nu}^{\rm beam}\lambda^{3}\
      \rm e^{\frac{1.44\times10^{4}}{T_d\lambda}}}{\kappa_{\rm m}(\lambda)\theta^{2}}}{\frac{Z{\odot}}{\rm Z}}
\label{eq:h2}
\end{equation}
where $S_{\nu}^{\rm beam}$ is the flux density in Jy/beam, $\lambda$ the
wavelength in $\mu$m, $\rm Z/Z{\odot}$ is the metallicity relative to
the solar metallicity (we assume $\rm Z/Z{\odot} = 1$) and $\theta$ is
the FWHM of the 100~m beam in arcseconds.  Here we assume that the dust temperature $T_d$ is
approximately equal to the gas temperature. We derive an effective
$\rm H_2$ column density of the order of 6 -- $20\times10^{22}$
cm$^{-2}$ toward the peaks of the dust emission. Together with the
\AMM\ column density, this is used to determine the ammonia abundance
($\rm \chi_{\rm NH_3}=N(NH_3)/N(H_2)$ ). We determine the peak
effective $\rm H$ column density from the SCUBA map and the \AMM\
column density for the same position for each clump and get an upper
limit for the fractional \AMM\ abundance as listed in
Table~\ref{tab:physical_parameters} . The average abundance
is $\sim 4\times10^{-8}$, however has a low value towards  G79.27+0.38 \AMM\ P2 
($7\times10^{-9}$, as listed in
Table~\ref{tab:physical_parameters}).

The mass of the each cloud can be determined from the dust continuum and
from \AMM\ emission. The gas mass is derived from the dust using
the relation\citep{launhardt1997:mass}
 \begin{equation}
M(\rm H_{2})[{M_\odot}] =
1.8x10^{-14}{\frac{S_{\nu}\lambda^{3}d^{2}\rm e^{\frac{1.44\times10^{4}}{T_d\lambda}}}{\kappa_m(\lambda)}}{\frac{Z{\odot}}{\rm
Z}}[ M_{\odot}]
\end{equation}
where $S_{\nu}$ is the integrated flux density in Jy and the other parameters
are the same as in eq.(~\ref{eq:h2}).

Our estimates for the mass and \hh\ column density are in general agreement 
with the previous work of \citet{carey2000:irdc} for the submillimeter cores.

The total gas mass can be derived from the \AMM\ column density maps
assuming a uniform fractional abundance of the molecule, for those
sources without dust continuum data.  The gas mass derived from SCUBA
observations can be directly compared with the virial mass
estimate. The virial parameter \citep{bertoldi1992:pr_conf_cores} for a clump
is defined as
\begin{equation}
     \alpha = {\frac{{5}{\sigma^{2}}{\rm R}}{\rm GM}}
\label{eq:alpha}
\end{equation}
where $\sigma$ is the three dimensional root mean square (r.m.s)
velocity dispersion and R is the radius of the clump, and M is the gas
mass. Note that $\sigma = \sqrt{3/8\rm{ln}2}\rm\times{FWHM}$.
The virial mass is defined as
$M_{\rm vir} = 5 \sigma^{2}{\rm R}/\rm {G}$. For the
clumps to be stable against collapse, $\alpha\sim1$. We find an
average value of $2.1$ for the virial parameter toward the individual cores. As the
uncertainty in the dust opacity alone is a factor of 2, $\alpha\sim1$, and
most of these cores appear to be virialised. Therefore, the core structure is
consistent with them being supported by turbulent pressure without any
evidence of external bounding pressures.  
G28.34+0.06 \AMM\ P4 is an extreme case, where $\alpha = 6.53$.

 \section{Discussion \label{sec:disc}}
\subsection{Gas Temperature and Distribution}

In G11.11$-$0.12, the position \AMM\ P2 is a faint MSX source with no
counterpart in the visible or NIR; it is very likely that this is a
heavily embedded protostar \citep{pillai2005a:g11}. The southern
clumps are colder by $\sim2$ -- 3~K than the northern clumps. Thus,
these southern filaments may be at an earlier stage of evolution than the
cores belonging to the northern filament, where star formation
activity might heat up the gas. The core to core variation
in temperature within an IRDC is observed toward several other sources (see \S\ref{subsec:tkin}). Therefore, these dense cores in IRDCs are ideal sites
for investigating the initial conditions in forming a massive star.

The gas temperatures we derive generally agree with the
dust temperatures of \citet{carey2000:irdc} based on submm observations and an
assumed dust emissivity index $\beta=1.75$. $T_{\rm gas}$ might
be lower limits for those sources that show up as bright and
compact objects in the submm dust emission, indicating the presence of
a heavily embedded object, but reasonable for the cold gas envelope.

\subsection{Physical Properties, Chemistry \& Lifetime}

The excellent correlation between the \AMM\ line and submm continuum emission is
consistent with the predictions of chemical models that \AMM\ is
relatively more abundant in high density region than other molecules
(\citealt{bergin2002:chemistry}). The column densities translate to
extremely high extinction values of $55-450$~mag. The discrepancies in
\AMM\ and \ALD\ linewidths reported by \citet{carey1998:irdc} (see
\S\ref{subsec:linewidths}) indicate the differences in the gas volume
traced by the two molecules. The line wings seen in \ALD\ and CH$_3$OH
(Leurini et al.\ 2006 submitted) gas toward some sources is most likely
high-velocity outflow. The critical density of the \ALD\ lines
studied (10$^{6}$~\percc) is much higher than that of the \AMM\
inversion lines (10$^4$~\percc). Hence, while physical parameters
derived from \AMM\ are representative of the general, cool IRDC
material, \ALD\ probes the dense gas, which might be influenced by
embedded protostars. The derived \AMM\ abundance of 0.7 --
$10.1\times10^{-8}$ together with the centrally condensed \AMM\
emission is consistent with the chemical model predicted for
pre-protostellar cores \citep*{bergin1997:chemistry}. These values
imply that \AMM\ is overabundant by factors of 5 -- 10 relative to
"normal"(= lower density) and less turbulent dark clouds. In contrast,
H$_2$CO is under-abundant by a factor of $\sim 50$
\citep{carey1998:irdc}. Hence the IRDCs appear to exhibit a complex
chemistry.

 We derive the flow crossing time, for the
cloud to disperse due to its own internal motions.  Perturbations
within the cloud would disintegrate the cloud unless the propagation
wavelength is much smaller than the distance across the cloud which it
traverses. The estimated dynamical timescales of $\sim 0.5$ -- 2~Myrs,
provide an upper limit to the life time of these
clouds.
The virial masses for the different clumps span a wide range from 50 to 
$7880~M_{\odot}$. As typical for intermediate to high mass star-forming 
regions, most of the masses are skewed to values
$>500~M_{\odot}$. The virial parameter $\sim1$ for most of the clumps
which are stable against collapse.

\subsection{IRDCs and Other Star-forming Regions}

In a comparison of our \AMM\ results with the potential low mass
counterpart B68 \citep{lai2003:b68}, we find that the linewidths and the
virial masses are much higher towards our cores. The sizes and the
column density and most importantly the masses we derive are also
significantly higher, thus making IRDCs potential candidates as sites
of high-mass star formation. The (1,1) and (2,2) linewidths for IRDCS
are slightly lower but the \AMM\ column densities are comparable to
the massive dense cloud identified in \AMM\ in NGC6334 I(N) by
\citet{kuiper1995:ngc6334}. In terms of masses and sizes and
temperatures, our sources are similar to those of
\citet{garay2004:new_cores}. Star formation probably has already
started in some of them.

In Table~\ref{tab:nh3_lowmass}, we present the mean
properties of the dense cores of our sample with those of the
cores in Taurus, Perseus and the Orion A complex as given in
\citet{ladd1994:nh3_perseus}. It is evident that IRDC cores are on
average highly turbulent, larger with much higher masses than the other cores
while having similar temperature as the other cores.
The IRDC masses derived here are for an \AMM\ abundance of
$10^{-7}$ as used by \citet{ladd1994:nh3_perseus}. But this is almost
an order of magnitude higher than what we derive and hence the mean
mass quoted in Table~\ref{tab:nh3_lowmass} will be an order
of magnitude higher, several 1000 \Msol\ instead of several 100.
Compared to local dark clouds, IRDCs pile up significantly large amounts of 
mass and have supersonic internal motions.  But how much of
this mass goes into forming stars of low, intermediate or high mass
is yet to be answered.
\citet{tan2005:cluster_iau} find that a group of
local IRDCs have masses of a few $10^3$ -- $10^4$~\Msol and mean surface
density of $\Sigma \sim 0.1$~g$\rm cm^{-2}$. They find that this is 3 times
the mean surface density of a Giant Molecular Cloud (GMC) and very similar to
the values found in more evolved systems like the Orion Nebula Cluster (ONC).
Subsequently, they suggest that IRDCs forming from GMCs are the initial
conditions for star clusters.

\citet{tan2005:cluster_iau} define $\Sigma$ as $\Sigma=M/(\pi R^2)$,
and if we compute the mean surface density for our sample and the
local dark clouds (the mean Mass and the size as given in
Table~\ref{tab:nh3_lowmass}, we arrive at values between 0.08 --
0.4. If the mean surface density were to be a measure of star
formation efficiency, then IRDCs have a higher value compared to
Taurus but not very much  higher than Perseus which is understood to be
an intermediate star forming region.

Studies of relatively local cluster forming regions like the $\rho$
Ophiuchus cloud, the Serpens molecular cloud and Orion B molecular cloud
(\citealt{motte1998:ophiuchus}, \citealt{testi1998:serpens},
\citealt{johnstone2001:orion}) find that the mass spectrum of their cores
are very similar to that of the stellar IMF. If this is indeed true,
then the fraction of the core mass going into forming stars would be
independent of mass and the stellar IMF would mainly be determined by the
cloud fragmentation process \citep{blitz1999:molecular_clouds}.  If
we assume a star formation efficiency of 30\% in an IRDC core of mean mass
$\sim 500$~\Msol
and adopt
the standard IMF (with power law indices $2.3 \pm 0.3$, $1.3 \pm 0.5$
and $0.3 \pm 0.7$ for masses $> 0.5$, 0.08 -- 0.5 and $< 0.05$~\Msol\
respectively), then $\sim 116$ stars could form in the core.  Out of
this, $\sim 2 $ stars could be of high-mass stars ($\ge 8$~\Msol)
while $64$ would be low-mass/intermediate stars ($0.5
\le$~M/\Msol~$\le8$) while the rest would be very low mass and sub-stellar objects.  Indeed
there is growing evidence of star formation in these cores
(\citealt{rathborne2005:irdc_astroph}, \citealt{pillai2005a:g11},
\citealt{ormel2005:irdc}).

 \subsection{IRDCs Formed by Interaction with SNRs ?}

From the velocity structure, G11.11$-$0.12 \AMM\ P2 appears blue
shifted while the clump \AMM\ P1 towards the north and \AMM\ P4
towards the south appears redshifted relative to the LSR
velocity. This is thus not a case of smooth velocity gradient along
the filament. It might be possible to explain the observed velocity
structure, if we assume that the entire filament seen in projection is
part of a unbound system where P1 and P4 lie at the same distance
along the line-of-sight (l.o.s) at diametrically opposite ends while
P2 is further in the front. A massive wind-driven process which might
have taken place in the close vicinity of the cloud could explain such
a structure. Recent wide field MIR images of this region released
from Spitzer show that this cloud has a filamentary concave structure
spread over several parsecs with a significant density enhancement in
the center \citep*{menten2005:iau}.  The morphology bears
remarkable similarity to the structures predicted by 3-D numerical
calculations simulating the impact of a planar shock front on an
isolated globule \citep{boss1995:shock}. The best known case 
of a ongoing SN-cloud interaction is in IC 443, where a 10$^4$ yr old SNR in the
GEM OB1 association with a shock speed of 40~\kms impacts the cloud
resulting in highly excited molecular gas with very broad linewidths. 
However, a weaker shock (likely from a more evolved SNR) would
result in temporary distortion and compression of the molecular cloud,
followed by rebound to a equilibrium.  An evolved SNR at an age of 10$^5$ yr
with a shock speed of 100~\kms, is one of the three likely weak
shock waves that \citet{boss1995:shock} propose to influence a cloud
without destroying it. 
Such a shock wave would have already
traversed 25~pc. \citet{brogan2004:snr} recently discovered the
supernova remnant G11.03$-$0.05 which is within $7'$ of the
cloud. They claim that the SNR is not young.

The SNR has a
shell-like appearance with a diameter of  $\sim 8$~\arcmin\ and very weak 
emission at centimeter wavelengths. \citet{brogan2004:snr} find that a pulsar
PSRJ1809$-$1917 is located about $8.5'$ from the SNR and is probably
associated with the SNR.  
Considering the respective uncertainties of their distance estimates, the 
pulsar ($4 \pm 1$~kpc) and G11.11-0.12 ($3.6 \pm 0.5$~kpc) appear at or near 
the same distance. 

In G33.71-0.01, we find another where the cloud seems to
have undergone a shock at the centre (see Fig.~\ref{fig:g33pv}). Note
that there is a large linewidth towards the central core of this
filament. The (0,0) velocity is also different from the cores toward
the south and north segment of the filament. The SNR G33.6+0.1 is
within $5'$ of this core and previous observations by
\citet{green1989:snr} shows an unusually broad OH absorption feature near
105~\kms\ towards this SNR. The $\rm{HCO^{+}}$ and $^{12} \rm CO~{\rm
J=1-0}$ observations by \citet{green1992:g33snr} reveal material
shocked by the interaction of the SNR with the adjacent molecular
cloud. The 2$\sigma$ detection of a large linewidth feature $\Delta v
\ge 8$~\kms\ centered at $105$~\kms\ in \AMM\ (1,1) and (2,2) might also
be a sign of interaction of the cloud with the SNR.

All IRDCs except those in the Cygnus-X region have a SNR in their
vicinity, although an association, which would require observations
with shock tracers, cannot be verified yet.
To estimate the chance occurance of an IRDC and SNR on the sky, we sample
$\sim$17.6 square degrees in which \citep{green2002:snr_catalogue} and
\citet{brogan2004:snr} have identified a total of 22 remnants.  The likelihood
of a chance positional occurance within 7\arcmin\ of an IRDC is small ($\le
0.05$) given a SNR surface density of 1.25 remnants per square degree.  It is
very likely that the SNRs are physically associated with the IRDCs G11.11 and
G33.71-0.01.

\subsection{IRDCs in the Framework of an Evolutionary Sequence \label{subsec:evolution} }
\label{fig:12}
\begin{figure}
\begin{center}
\includegraphics[height=1.5\linewidth, angle=-0]{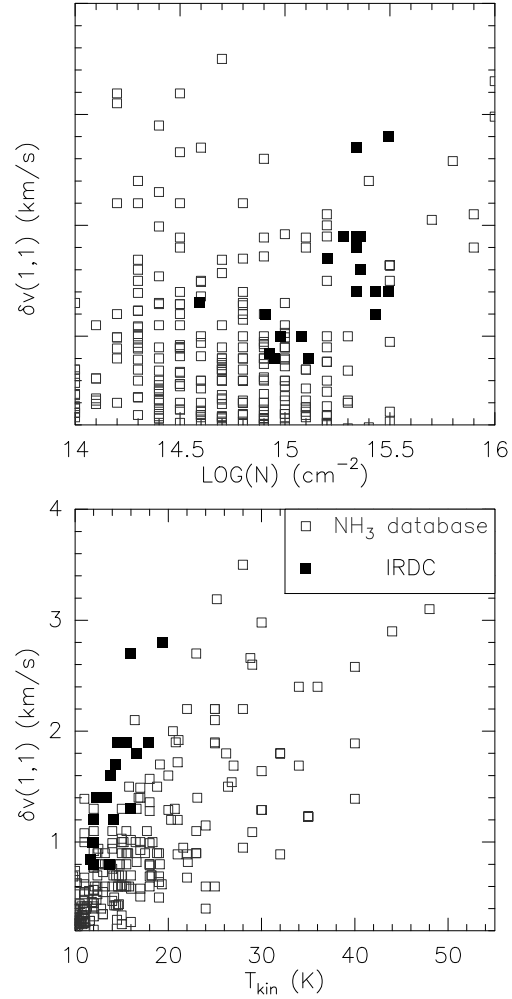}
\end{center}
\caption{\textit{Bottom panel:}The correlation plot of kinetic temperature with the linewidths.
The black filled squares indicate the IRDCs and the grey unfilled
squares the cores from the \AMM\ database. IRDCs have larger velocity
dispersions while the rotational temperatures are comparable.
\textit{Top panel:}The correlation plot of \AMM\ column density with
the linewidths shows that the average column density of IRDCs are
higher.\label{fig:nh3jijina}}
\end{figure}

To obtain general statistics on IRDCs we compare in
Fig.~\ref{fig:nh3jijina} the core gas properties of our sample of 9
sources with the cores presented in the \citet{jijina1999:database} \AMM\
database. The database consists of 264 dense cores, with and without
associations with young stellar objects. For the large linewidths of
IRDCs ($\Delta \rm v > 0.8$~\kms), there is a distinct trend for the
IRDCs to be colder relative to the cores from the database. The \AMM\
column density is also higher for the IRDCs, surpassed only by a few
high column densities from low mass cores observed with the high
angular resolution of the VLA, which evidently sees more of the core
interiors that  are possibly associated with YSOs.

\begin{figure}
\begin{center}
\includegraphics[height=1.5\linewidth, angle=0]{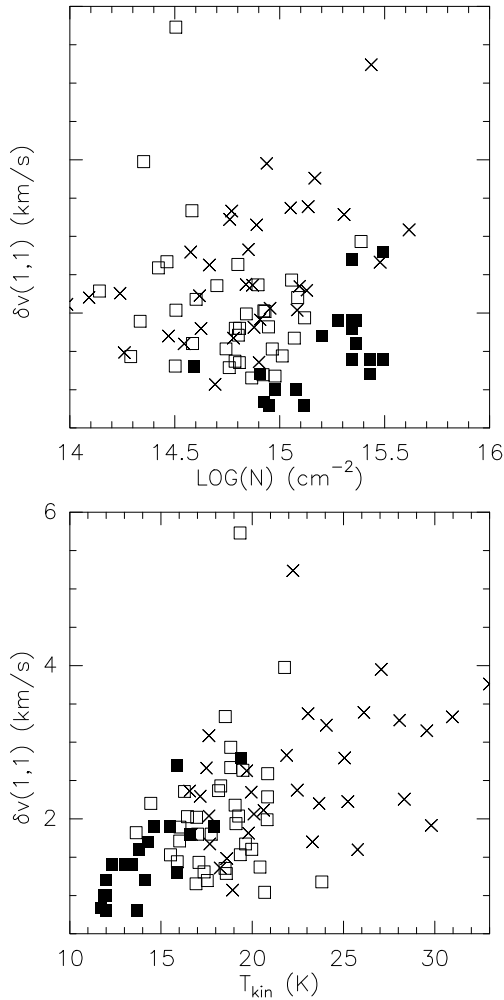}
\end{center}
\caption{\textit{Bottom panel:}The correlation plot of kinetic
temperature with the linewidth. The black filled squares indicate the
IRDCs,the crosses the sources from the Wood and Churchwell catalogue
and the unfilled squares \citet{beuther2002:dust_cs} sources. IRDCs
are colder and on average have lower linewidths.\textit{Top
panel:}The correlation plot of \AMM\ column density with the linewidths.
 \label{fig:nh3beuther}}
\end{figure}

In Fig.~\ref{fig:nh3beuther} we present a similar comparison with a
sample of UCH{\sc ii} regions from the \citet{wood1989:uchii}
catalogue and the sources from the \citet{beuther2002:dust_cs} study
of high-mass protostellar objects (HMPOs).
While recent studies reveal that HMPOs are in a pre-UCH{\sc ii} region
phase (\citealt{molinari2002:hmpo}, \citealt{beuther2002:dust_cs}),
the nature of the stage preceding HMPOs has not been studied in
detail. IRDCs are ideal candidates for this pre-HMPO stage. In
Fig.~\ref{fig:nh3beuther} we present properties of source samples
believed to cover the earliest phases of massive star formation based
on \AMM\ observations. There is no clear trend in linewidths,
however IRDCs have a significantly lower average linewidth ($\Delta \rm
\overline{v} = 1.51$~\kms) 
than HMPOs ($\Delta \rm \overline{v} = 2.05$~\kms) and UCH{\sc
ii} regions ($\Delta \rm \overline{v} = 2.52$~\kms).  There is a
distinct temperature trend from the low temperatures of the IRDCs
($\rm \overline{T} = 13.9$~K) to increasing temperatures 
for the IRAS selected high-mass objects ($\rm \overline{T}
= 18.34$~K) and high temperatures for the objects associated with
UCH{\sc ii} regions ($\rm \overline{T} = 22.6$~K). Temperature and
linewidths must be understood as averages over the core and parts of
the envelope since the beam is 40\arcsec.

There is no clear trend in column density. However, on average the
column densities of IRDCs are high compared to the other two samples.
We interpret the clear trend in temperature and the tentative
differences in \AMM\ linewidths and column densities (N[NH$_3$]) 
as possible manifestation of an evolutionary
sequence.  Starless cores on the verge of star formation are expected
to be cold (T$\le$20~K), to have high column densities and smaller
linewidths. The temperature and the linewidth in a core will increase
after the formation of an embedded protostar (via radiative heating
and injection of turbulence through outflows and winds), while the
envelope will be dispersed via outflows and winds and thus column
densities will decrease with time.

One would expect the highest ammonia column densities for the hot
sources due to evaporation of ammonia from the grains although in some
cases IRDCs have higher column densities. However, this effect might
be very localised to the hot cores (0.1 pc and smaller) and not be
true for the larger scale emission. Additionally some molecular
material might already be dispersed in the hot sources by the
interaction of the young OB clusters with their environment.
\label{fig:13}

 \section{Conclusions}
In this paper, we discussed ammonia observations of a selected set of
IRDCs and the derived physical properties.  Our results are as
summarised below.

The ammonia emission correlates very well with MIR absorption and
ammonia peaks distinctly coincide with dust continuum peaks. Several
cores are detected within the clouds with deconvolved sizes
smaller than the 40$''$ FWHM beam size.  We can constrain the
average gas temperature to between 10 and 20 K.

We observe high linewidths ($1 \le \Delta v/\rm km~s^{-1} \le 3.5$),
hence turbulence plays an important role in the stability of an IRDC.
There are significant velocity gradients observed between the
cores. The effect of external shock/outflow tracers , on the gas
kinematics is suggestive in some cases, but needs to be investigated
further.

The column densities translate to extremely high $A_V$ values (55 --
$450$~mag), therefore any active star formation would be heavily
embedded.  The total cloud gas mass derived from the \AMM\
data ranges from 10$^3$ -- 10$^4$ \Msol.  The virial parameter is
$\sim 1$ for most of the clumps, and the cores appear to be stable
against gravitational collapse.  As a result IRDCs are potential sites for
star formation.  If we were to adopt the stellar IMF and a star formation
efficiency of 30\%, then every IRDC
core could fragment to form $>100$ stars, with at least two high mass
stars ($>8$~\Msol).

The fractional abundance of \AMM\ (relative to $\rm H_2$) is 0.7 --
$10.0\times 10^{-8}$. This, together with the excellent correlation in
morphology of the dust and gas, is consistent with the time dependent
chemical model for \AMM\ of \citet{bergin1997:chemistry} and implies
that \AMM\ remains undepleted. The derived abundance is a factor 5 --
10 larger than that observed in local dark clouds while \ALD\ is
underabundant by a factor of $\sim 50$. Hence, the chemistry governing
these IRDCs might be complex and could be different from other parts of the
dense ISM.

The time scales
we derive for the clouds to disperse due to their own internal motions
of a few Myrs provides an upper limit to the life time of these
clouds.  We suggest that SNRs might be the trigerring mechanism
responsible for the formation of an IRDC.

The comparison of the physical properties from ammonia of our IRDCs
sample with other source samples -- HMPO's and UCH{\sc ii}s strongly
suggests that most of these IRDCs are the most likely candidates for
pre-protostellar cores of massive star formation.

\begin{table*}
\begin{center}
\caption{List of IRDCs Observed in \AMM (1,1) and (2,2)}.
\vspace*{2mm}
\begin{tabular}{llldd}
\hline
\hline
Name   &        R.A.(J2000)  &     Dec.(J2000)     &  \dlabel{$V_{\rm LSR}$~[\kms]}  &       \dlabel{D~[kpc]}   \\
\hline

G11.11-0.12 P1 & 18:10:29.27 & --19:22:40.3 & 29.2 &  3.6  \\
G19.30+0.07 & 18:25:56.78 & --12:04:25.0   &  26.3 &  2.2  \\
G24.72-0.75 & 18:36:21.07 & --07:41:37.7 & 56.4 & 1   \\
G24.63+0.15 & 18:35:40.44 & --07:18:42.3 & 54.2 & 3.6  \\
G28.34+0.06 P1 & 18:42:50.9 & --04:03:14 & 78.4 & 4.8 \\
G28.34+0.06 P2 & 18:42:52.4 & --03:59:54 & 78.4 & 4.8 \\
G33.71-0.01 & 18:52:53.81 & +00:41:06.4 & 104.2 & 7.2 \\
G79.27+0.38 & 20:31:59.61 & +40:18:26.4 & 1.2  & 1\\
G79.34+0.33 & 20:32:21.803 & +40:20:08.00 & 0.1 & 1\\
G81.50+0.14 & 20:40:08.29 & +41:56:26.4 & 8.7 & 1.3\\
\hline

\end{tabular}
\label{tab:source_list}
\end{center}
Notes: Columns are name, right ascension, declination, LSR velocity, distance. Positions and Distances are taken from \citet{carey1998:irdc}.
    The coordinates correspond to the reference positions of the maps.

\end{table*}
\begin{sidewaystable*}
\begin{minipage}{\textheight}
\caption{\AMM\ (1,1) and (2,2) Map Results: Peak Position.}.
\label{tab:line_parameters}
\centering
\begin{tabular}{rclllll}
\hline\hline
Source                    & Transition & $V_{\rm LSR}$   &  $T_{\rm MB}$         &  FWHM              &  $\tau_{\rm main}$ &  1~$\sigma$  \\
                       &           &  [\kms]            &  [K]                  &  [\kms]              &                   &   [K~\kms]    \\
\hline
G11.11$-$0.12 \AMM\ P1    & 1--1 &30.44   ($\pm$0.02) & 5.47  ($\pm$0.96) & 1.27  ($\pm$0.06) & 3.52  ($\pm$0.39) &0.42 \\
                          & 2--2 &30.42   ($\pm$0.09) & 1.97  ($\pm$0.32) & 1.56 ($\pm$0.30) &          &     \\
G19.30$+$0.07 \AMM\ P1    & 1--1 &26.45   ($\pm$0.03) & 5.17  ($\pm$0.8)  & 1.88  ($\pm$0.08) & 1.88  ($\pm$0.26) &0.53 \\
                          & 2--2 &26.64   ($\pm$0.06) & 2.86  ($\pm$0.27) & 1.57   ($\pm$0.17) &          &     \\
G24.72$-$0.75 \AMM\ P1    & 1--1 &58.37   ($\pm$0.07) & 2.76  ($\pm$0.45) & 2.57  ($\pm$0.12) & 2.81  ($\pm$0.47) &0.25 \\
                          & 2--2 &58.47   ($\pm$0.12) & 1.99  ($\pm$0.23) & 2.97  ($\pm$0.24) &          &     \\
G24.63$+$0.15             & 1--1 &53.11   ($\pm$0.04) & 2.40   ($\pm$0.34) & 1.7   ($\pm$0.09) & 2.36  ($\pm$0.43)  &0.16\\
                          & 2--2 &52.89   ($\pm$0.09) & 0.98  ($\pm$0.12) & 2.42 ($\pm$0.24) &          &     \\
G28.34$+$0.06 \AMM\ P1    & 1--1 &79.14   ($\pm$0.02) & 2.88  ($\pm$0.42) & 2.67  ($\pm$0.04) & 2.00  (0.11)  &0.16\\
                          & 2--2 &79.15   ($\pm$0.04) & 1.37  ($\pm$0.05) & 3.29  ($\pm$0.11) &           & \\
G33.71$-$0.01 \AMM\ P1    & 1--1 &104.31  ($\pm$0.09) & 2.52  ($\pm$0.43) & 2.64  ($\pm$0.2)  & 1.5   ($\pm$0.52)  &0.12\\
                          & 2--2 &104.4   ($\pm$0.17) & 1.14  ($\pm$0.22) & 2.40 ($\pm$0.35) &           & \\
G79.27$+$0.38 \AMM\ P1    & 1--1 &1.65    ($\pm$0.02) & 2.28  ($\pm$0.38) & 1.04  ($\pm$0.05) & 2.49  ($\pm$0.27)  &0.31\\
                          & 2--2 &1.63    ($\pm$0.05) & 0.61  ($\pm$0.08) & 0.98  ($\pm$0.14) &           & \\
G79.34$+$0.33 \AMM\ P1    & 1--1 &0.31    ($\pm$0.01) & 2.53  ($\pm$0.30)  & 1.23  ($\pm$0.03) & 1.45  ($\pm$0.13)  &0.19\\
                          & 2--2 &0.25    ($\pm$0.04) & 0.79  ($\pm$0.07) & 1.29  ($\pm$0.13) &           & \\
G81.50+0.14               & 1--1 &-4.53   ($\pm$0.05) & 2.21  ($\pm$0.35) & 1.35  ($\pm$0.15) & 0.37  ($\pm$0.46) &0.27 \\
                          & 2--2 &-4.69   ($\pm$0.12) & 0.67  ($\pm$0.21) & 1.00  ($\pm$0.34) &              & \\

\hline
\hline
\end{tabular}

\vfill
Notes: Columns are name, \AMM\ (J,K) transition, LSR velocity, (1,1)
main beam brightness temperature, full linewidth at half maximum, main
group optical depth and the 1~sigma noise level in the \AMM\
(1,1) integrated intensity map (\ref{fig:msx_nh3}). The error estimates
are given in brackets.
\end{minipage}
\end{sidewaystable*}

\begin{sidewaystable*}
\begin{minipage}{\textheight}
\caption{Physical Properties of Observed IRDCs.}
\centering
{\scriptsize
\begin{tabular}{ccrrrrcdddddc}
\hline
\hline
Source                        & Offsets       & $T_{\rm EX}$     & $T_{\rm ROT}$   & $T_{\rm KIN}$ & $N({\rm NH_3})$ & $N({\rm H+H_2})$ & \dlabel{$\chi_{\rm NH_3}$}  & \dlabel{Mass}  & \dlabel{Size} & \dlabel{$M_{\rm vir}$} & \dlabel{$\alpha$} & \dlabel{$M_{\rm NH_3 }$} \\

                              & [arcsecs]   & [K]    & [K] & [K]    & [$10^{14}~{\rm cm^-2}$]   &  \dlabel{[$10^{22}~{\rm cm^-2}$]} &  \dlabel{[$10^{-8}$]}   &\dlabel{$M_{\odot}$}    & \dlabel{[arcsecs]}& \dlabel{[$M_{\odot}$]} & \dlabel{}& [$10^4$~$M_{\odot}$] \\
\hline
G11.11$-$0.12  \AMM\ P1       & (80,40)       &   7.6 ($\pm$0.7) & 12.3 ($\pm$1.2) & 12.7 & 41.5 ($\pm$ 6.4)  &  8.4 & 4.9  & 485 & 27 & 468   & 1.0    & 18.1 \\
G11.11$-$0.12 \AMM\ P2        & (0,0)         &   6.6 ($\pm$0.4) & 13.4 ($\pm$1.0) & 13.8 & 30.5 ($\pm$ 3.3)  & 10.2 & 2.9  & 172 & 13 & 268   & 1.6    &   \\
G11.11$-$0.12 \AMM\ P3        & (-300,-325)   &   7.1 ($\pm$0.3) & 14.1 ($\pm$9.7) & 14.5 & 60.4 ($\pm$ 11.4) &  8.7 & 6.9  & 647 & 32 & 860   & 1.3    &    \\
G11.11$-$0.12 \AMM\ P4        & (-300,-400)   &   5.7 ($\pm$0.5) & 10.5 ($\pm$1.0) & 10.8 & 50.9 ($\pm$ 10.6) & 16.1 & 3.2  & 788 & 25 & 555   & 0.7    &      \\
G19.30$+$0.07 \AMM\ P1$^{n}$  & (20,20)       &   8.2 ($\pm$0.9) & 17.6 ($\pm$2.0) & 18.4 & 25.9 ($\pm$ 3.2)  &      &     &     & 38 & 893   &        & 0.9 \\
G19.30$+$0.07 \AMM\ P2 $^{n}$ & (-60,-20)     &   5.9 ($\pm$0.7) & 13.8 ($\pm$1.8) & 14.3 & 33.3 ($\pm$ 6.9)  &      &     &     & 45 & 823   &        &        \\
G24.72$-$0.17 \AMM\ P1 $^{n}$ & (150,220)     &   6.6 ($\pm$0.6) & 17.4 ($\pm$2.2) & 18.2 & 44.3 ($\pm$ 5.7)  &      &     &     & 34 & 2450  &        & 1.4 \\
G24.63$+$0.15$^{n}$           & (0,0)         &   5.0 ($\pm$0.5) & 14.3 ($\pm$1.4) & 14.8 & 24.1 ($\pm$ 4.4)  &      &     &     & 43 & 1358  &        & 0.7 \\
G28.34$+$0.06 \AMM\ P1        & (0,0)         &   5.7 ($\pm$0.2) & 16.0 ($\pm$1.2) & 16.6 & 32.1 ($\pm$ 2.2)  &  5.8 & 5.5  &904  & 22 & 2280  & 2.5    &  25.4 \\
G28.34$+$0.06 \AMM\ P2        & (20,200)      &   5.5 ($\pm$0.2) & 15.4 ($\pm$1.4) & 16.0 & 38.6 ($\pm$ 3.2)  & 32.7 & 1.2  &2310 & 45 & 2928  & 1.3    &  \\
G28.34$+$0.06 \AMM\ P3        & (-60,-60)     &   4.8 ($\pm$0.2) & 12.8 ($\pm$1.2) & 13.2 & 47.2 ($\pm$ 6.1)  &  5.9 & 8    &374  & 21 & 1398  & 3.7    &      \\
G28.34$+$0.06 \AMM\ P4        & (-30,50)      &   4.8 ($\pm$0.2) & 15.9 ($\pm$2.2) & 16.4 & 37.4 ($\pm$ 4.9)  &  3.7 & 10.1 &147  & 16 & 960   & 6.5    &   \\
G28.34$+$0.06 \AMM\ P5        & (50,40)       &   5.6 ($\pm$0.3) & 13.6 ($\pm$1.3) & 14.1 & 29.6 ($\pm$ 3.8)  &  6.3 & 4.7  &528  & 24 & 1378  & 2.6    &     \\
G33.71$-$0.01 \AMM\ P1 $^{n}$ & (-40,-200)    &   5.6 ($\pm$1.1) & 16.6 ($\pm$2.3) & 17.2 & 23.0 ($\pm$ 6.6)  &      &     &     & 52 & 7880  &        & 15.6 \\
G33.71$-$0.01 \AMM\ P2 $^{n}$ & (60,100)      &   5.3 ($\pm$0.3) & 17.1 ($\pm$2.0) & 17.8 & 29.0 ($\pm$ 3.3)  &      &     &     & 60 & 4928  &        &      \\
G79.27$+$0.38 \AMM\ P1        & (-20,0)       &   4.8 ($\pm$0.2) & 12.4 ($\pm$0.9) & 12.8 & 20.5 ($\pm$ 2.5)  & 14.2 & 1.4  & 84  & 32 & 155   & 1.8    & 0.4 \\
G79.27$+$0.38 \AMM\ P2        & (-140,40)     &   6.0 ($\pm$0.7) & 11.2 ($\pm$1.4) & 11.5 & 11.3 ($\pm$ 2.7)  & 12.1 & 0.9  & 55  & 25 & 53    & 1.0    &     \\
G79.27$+$0.38 \AMM\ P3        & (-250,70)     &   3.3 ($\pm$2.2) & 11.7 ($\pm$1.6) & 12.0 & 10.0 ($\pm$ 25.0) &  6.5 & 1.5  & 19  & 20 & 40    & 2.1    &         \\
G79.34$+$0.33 \AMM\ P1        & (0,0)         &   5.8 ($\pm$0.3) & 14.1 ($\pm$0.8) & 14.6 & 11.9 ($\pm$ 1.0)  & 17.5 & 0.7  & 143 & 39 & 178   & 1.2    & 0.2 \\
G81.50$+$0.14 $^{n}$          & (260,-375)    &   5.6 ($\pm$1)   & 14.9 ($\pm$1.3) & 15.4 &  7.6 ($\pm$ 2.0)  &      &     &     & 78 & 570   &       & 0.1 \\
\hline
\end{tabular}
\label{tab:physical_parameters}
\vfill

Notes: Columns are name, Offset Position from reference, excitation temperature, \AMM\ rotational temperature, kinetic temperature, \AMM\ column density,  $\rm H_2$ column density, \AMM\ abundance, gas mass estimated from SCUBA data, size of the cores, virial masses from \AMM\, the alpha parameter defined as in total cloud mass from \AMM\. The formal errors are given in brackets. $^{n}$ are those sources for which the sizes
have been derived from \AMM\ data. The $\rm H_2$ column densities are estimated from 850 $\mu$m SCUBA map \citep{carey2000:irdc}).}
\end{minipage}
\end{sidewaystable*}

\begin{table*}
\begin{center}
\caption{Mean Values of \AMM\ Core Samples.}\label{tab:nh3_lowmass}
\vspace*{2mm}
\begin{tabular}{rddddd}
\hline
\hline
Sample & \dlabel{FWHM (\kms)} & \dlabel{$T_{\rm kin}$(K)} & \dlabel{size(pc)}& \dlabel{$M_{\rm NH_3}$ ($M_{\odot}$)}& \dlabel{$\Sigma$} \\
\hline
Taurus&0.33  & 10& 0.06 & 1.1 & 0.08 \\
Perseus&0.55  &13 & 0.12 &  9 & 0.16 \\
Orion&1.12 & 17& 0.15 & 21 & 0.24 \\
IRDC cores& 1.7 &15& 0.57& 492 & 0.40 \\
\hline
\end{tabular}
\end{center}
Notes: Columns are the source sample, the mean FWHM of the \AMM\ (1,1)
line, kinetic temperature, size of the core, mass as determined from
\AMM\ assuming a fractional abundance of \AMM\ relative to $\rm H_2$
and the mean surface density.  The mean values for dense cores in
Taurus, Perseus and Orion have been taken from \citet{ladd1994:nh3_perseus}.

\end{table*}

\acknowledgements{ This research has made use of the NASA/ IPAC
Infrared Science Archive, which is operated by the Jet Propulsion
Laboratory, California Institute of Technology, under contract with
the National Aeronautics and Space Administration.  S.J.Carey acknowledges
support from a NASA Long Term Space Astrophysics grant.  T.Pillai was
supported for this research through a stipend from the International
Max Planck Research School (IMPRS) for Radio and Infrared Astronomy at
the University of Bonn.  T.Pillai thanks her colleague J.Kauffmann for
useful discussions and comments on the manuscript.
 }

\bibliographystyle{aa}
\bibliography{bib_astro}
\end{document}